\documentclass[preprint,review,12pt]{elsarticle}

\pdfoutput=1 

\usepackage{amssymb}
\usepackage{eucal}
\usepackage{amsopn}

\newtheorem{thm}{Theorem}

\newcommand{\scr}{\EuScript}
\DeclareMathSymbol{\upset}{\mathopen}{symbols}{"22}
\DeclareMathSymbol{\downset}{\mathopen}{symbols}{"23}

\newcommand{\myF}{\scr F}
\newcommand{\myG}{\scr G}

\newcommand{\nO}{\mathrm{O}}
\newcommand{\per}{\ensuremath{\operatorname{per}}}
\newcommand{\pertp}{\ensuremath{\mathrm{per}'\,}}

\newcommand{\binsum}[2]{{~#1 \choose \downset#2}}

\newcommand{\be}{\begin{eqnarray}}
\newcommand{\ee}{\end{eqnarray}}
\newcommand{\bes}{\begin{eqnarray*}}
\newcommand{\ees}{\end{eqnarray*}}

\begin{document}

\begin{frontmatter}

\title{On evaluation of permanents}

\author[lund]{Andreas Bj\"orklund}
\ead{andreas.bjorklund@yahoo.se}
\author[lund,copenhagen]{Thore Husfeldt}
\ead{thore.husfeldt@cs.lu.se}
\address[lund]{Lund University,
  Department of Computer Science,\\
  P.O.Box 118, SE-22100 Lund, Sweden}
\address[copenhagen]{IT University of Copenhagen,\\
  Rued Langgaards Vej 7, 2300, K\o{}benhavn S, Denmark}
\author[hiit]{Petteri Kaski\fnref{AF}}
\ead{petteri.kaski@cs.helsinki.fi}
\author[hiit]{Mikko Koivisto\fnref{AF}}
\ead{mikko.koivisto@cs.helsinki.fi}
\address[hiit]{Helsinki Institute for Information Technology HIIT,\\
  Department of Computer Science, University of Helsinki,\\
  P.O.Box 68, FI-00014 University of Helsinki, Finland}

\fntext[AF]{This research was supported in part by the Academy of Finland, 
Grants 117499 (P.K.) and 125637 (M.K.).}


\end{frontmatter}

The {\em permanent} of an $m\times n$ matrix $A = (a_{ij})$, 
with $m\leq n$, is defined as 
\bes
	\per A 
	\;\doteq\; \sum_{\sigma} a_{1\sigma(1)}\,a_{2\sigma(2)}\cdots
	a_{m\sigma(m)}\;, 
\ees
where the summation is over all injections $\sigma$ from $M \doteq
\{1,2,\ldots,m\}$ to $N\doteq \{1,2,\ldots,n\}$.  While studies on permanents --
since their introduction in 1812 by Binet \cite{Binet} and Cauchy \cite{Cauchy}
-- have focused on matrices over fields and commutative rings, we generally only
assume the entries are from some {\em semiring}, that is,  multiplication need
not commute and additive inverses need not exist. 

In this note, we give simple algorithms to evaluate the permanent of a given
matrix. In arbitrary semirings, we apply Bellman--Held--Karp type dynamic
programming \cite{Bellman,Bellman2,HeldKarp} across column subsets; in
commutative semirings, a ``transposed'' variant is shown to be considerably
faster. In arbitrary rings, the starting point is Ryser's classic algorithm
\cite{Ryser} that we manage to expedite for rectangular matrices, but that
remains the fastest known algorithm for square matrices;  again, in commutative
rings, a transposed variant is shown to be substantially faster for rectangular
matrices. 

To state our main results,  we take the {\em time} requirement of an algorithm
as the number of additions and multiplications it performs, while the  {\em
space} requirement is taken as the maximum number of semiring elements that it
needs to keep simultaneously in memory at any point in the computation.  Also, 
denote by $\binsum{q}{r}$ the sum of the binomial coefficients  ${q \choose 0} +
{q \choose 1} + \cdots + {q \choose r}$.

\begin{thm}\label{thm:main}
The permanent of any $m \times n$ matrix, $m \leq n$, 
can be computed
\begin{enumerate}
\item[(i)] in semirings in 
time $\nO\Big(m\binsum{n}{m}\Big)$ and space $\nO\Big(\binsum{n}{m}\Big)$; 
\item[(ii)] in commutative semirings in 
time $\nO\big(m (n - m + 1) 2^m\big)$ and\\ space $\nO\big((n - m + 1) 2^m\big)$; 
\item[(iii)] in rings in 
time $\nO\Big(m \binsum{n}{m/2}\Big)$ and space 
$\nO\Big(\binsum{n}{m/2}\Big)$; and 
\item[(iv)] in commutative rings in 
time $\nO((mn-m^2+n)2^m)$ and space $\nO(n)$. 
\end{enumerate}
\end{thm}

All previous works we are aware of on evaluation of permanents assume 
commutativity, besides perhaps what is implicit in Ryser's formula, see
(\ref{eq:ryser}) below. For commutative {\em rings}, our bounds improve upon the
state-of-the-art achieved in a series of works based on arguably more involved
techniques:  Using the Binet--Minc formulas \cite{Minc}, Kawabata and Tarui
\cite{KawabataTarui} presented an algorithm that runs in time $\nO(n 2^m +
3^m)$  and space $\nO(n 2^m)$. Recently,  Vassilevska and Williams
\cite{VassilevskaWilliams} took a different approach and obtained improved
bounds $\nO(mn^32^m)$ and $\nO(n^22^m)$, respectively.  Finally, by a yet
different, algebraic approach, Koutis and Williams \cite{KoutisWilliams} further
improved these bounds to $\mathrm{poly}(m, n) 2^m$ and  $\mathrm{poly}(m, n)$.
For commutative {\em semirings}, Vassilevska and Williams
\cite{VassilevskaWilliams} gave a Gurevich--Shelah \cite{GurevichShelah} type
recursive partitioning algorithm running in  time $\mathrm{poly}(m, n) 4^m$ and
space $\mathrm{poly}(m, n)$. Koutis and Williams \cite{KoutisWilliams} presented
bounds comparable to  Theorem~\ref{thm:main}(ii) using a dynamic programming
algorithm  similar to ours but in an algebraic guise.

We begin without any further assumptions about the semiring and adopt the standard  dynamic
programming treatment of sequencing problems. That is, the
algorithm tabulates intermediate results $\alpha(i, J)$ 
for sets $J \subseteq N$ of size $i$, 
given by the recurrence
\bes
	\alpha(0, \emptyset)  \;\doteq\; 1\;,\quad
	\alpha(i, J) 
	\;\doteq\; 
	\sum_{j \in J} \alpha(i-1, J\setminus\{j\})\, a_{ij}
	\quad \textrm{for } i=1,2,\ldots,m\;. 
\ees
Here $J$ corresponds to the image $\sigma(\{1,2,\ldots,i\})$ of the injection
$\sigma$, and   
it is easy to show that the permanent of $A$ is obtained as the sum of the
terms $\alpha(m, J)$ over all $J \subseteq N$ of size $m$. 
Straightforward analysis proves Theorem~\ref{thm:main}(i).

In commutative semirings, we may transpose the previous algorithm, as follows.
The idea is to go through the column indices $j$ one by one, associating $j$
with either one row index $i$ not already associated with some other column, or
associating $j$ with none of the rows. Formally, for all 
$I \subseteq M$ define recursively
\bes
	\alpha(\emptyset, 0) 
	 & \doteq & 1\;,\quad
	\alpha(I, 0) 
	 \;\doteq\; 0
	 \quad \textrm{for } I\neq\emptyset\;,\\[\medskipamount]
	\alpha(I, j)
	 & \doteq &
	\alpha(I, j-1) +
	\sum_{i \in I} \alpha(I\setminus\{i\}, j-1)\, a_{ij}
	\quad \textrm{for } j=1,2,\ldots,n\;. 
\ees
Here $I$ corresponds to the preimage $\sigma^{-1}(\{1,2,\ldots,j\})$ of the
injection $\sigma$.  
One can show easily by induction that $\alpha(I, j)$ equals the permanent of the
submatrix of $A$ consisting of the rows $I$ and columns $\{1,2,\ldots,j\}$; 
in particular, $\alpha(M, n) = \per A$. To obtain the bounds in 
Theorem~\ref{thm:main}(ii), it remains to observe 
that $\alpha(I, j)$ needs to be computed only if $|I|\leq j
\leq n-m+|I|$, and thus, the time and space requirements are 
$\nO(m (n-m+1)2^m)$ and $\nO((n-m+1) 2^m)$, respectively.

In rings, we start with Ryser's inclusion--exclusion formula. Denote by 
$a_{i X}$ the partial row sum of the entries $a_{ij}$ with $j \in X$.
Ryser \cite{Ryser} found that  
\be \label{eq:ryser}
	\per A 
	\;=\; 
	\sum_{X \subseteq N \atop |X| \leq m}\,
	(-1)^{m-|X|}\,{n-|X| \choose m-|X|}\,
	a_{1X}\,a_{2X}\cdots a_{mX}\;.
\ee
(While Ryser's original derivation is for fields, it immediately extends to arbitrary 
rings.) Visiting the sets $X$, for instance, in the lexicographical order, 
the terms $a_{iX}$ can be computed in an incremental fashion, each   
in constant amortized time. Thus the permanent can be 
evaluated in time $\nO\Big(m \binsum{n}{m}\Big)$ and space $\nO(m)$. 
For square matrices this remains the most efficient way to evaluate the
permanent.

But, when $m$ is much less than $n$ we can, in fact, do significantly better. 
For any subset of rows $I\subseteq M$ and any subset of columns $J\subseteq N$,
let $A_{IJ}$ denote the corresponding submatrix of $A$.  For simplicity, assume
$m$ is even, and denote $K \doteq \{1,2,\ldots,m/2\}$  and $L\doteq
\{m/2+1,m/2+2,\ldots,m\}$. Now, we may write   $\per A$ as the sum of
the products  $\per A_{KP}\, \per A_{LQ}$  over all {\em
disjoint} pairs of subsets $P, Q \subseteq N$ with $|P| = |Q| = m/2$. While
computing the sum over the ${n \choose m/2}{n-m/2 \choose m/2}$ such pairs $(P,
Q)$ may look inadvisable at first glance, the following observation changes the
picture. 

For a set family $\myF$, denote by $\downset{\myF}$ the family of sets in $\myF$
and their subsets.  

\begin{thm}[Bj\"{o}rklund et al.\ \cite{BHKK2009}, Kennes \cite{Kennes}]
Let $f$ and $g$ be two functions from the subsets of a finite set $U$ to a ring
$R$. Then,
\be\label{eq:ds}
	\sum_{S, T \subseteq U\atop S\cap T=\emptyset} f(S)g(T) 
	& \,= & 
	\sum_{X \subseteq U} (-1)^{|X|}
	\Big(\sum_{S\supseteq X} f(S)\Big)
	\Big(\sum_{T\supseteq X} g(T)\Big)\;.
\ee
Furthermore, if\/ $\myF$ and\/ $\myG$ are given families of subsets of $U$ such that
$f$ and $g$ vanish outside\/ $\myF$ and\/ $\myG$, respectively, then the sum\/
$(\ref{eq:ds})$ can be computed with\/ 
$\nO\big(\,|U|\,(\,|\downset{\myF}|+|\downset{\myG}|\,)\big)$
ring and set operations, and with a storage for\/
$\nO(|\downset{\myF}|+|\downset{\myG}|)$ ring elements.
\end{thm}

\sloppy
To apply this result, we first note that 
the cardinality of 
$\downset{\{P \subseteq N :|P| = m/2\}}$ is $\binsum{n}{m/2}$. 
Second, note that the permanent 
$\per A_{KP}$, for all $P \subseteq N$ of size $m/2$, can be computed 
in time $\nO\Big(m\binsum{n}{m/2}\Big)$ and space 
$\nO\Big(\binsum{n}{m/2}\Big)$; similarly for the permanents $\per A_{LQ}$. 
Combining these bounds yields 
Theorem~\ref{thm:main}(iii). We also note without proof that the space
requirement can be reduced to $\nO(m)$ at the cost of 
an extra factor of $3^{m/2}$ in the time requirement; the idea is the same
as what we have recently used to count paths and packings \cite{BHKK2009}.

Finally, in commutative rings we may transpose Ryser's formula in analogue to
the transposed dynamic programming algorithm for commutative semirings. To this
end, denote by $a_{X j}$ the partial column sum of the entries $a_{ij}$ with $i
\in X$.  Then we may write 
\bes
	\per A
	\;=\;  
	\sum_{X\subseteq M}(-1)^{m - |X|}
	\sum_{p} a_{X 1}^{p_1}\,a_{X 2}^{p_2}\cdots a_{X n}^{p_n}\;,
\ees
where the inner-most summation is over all binary sequences  $p=p_1 p_2 \cdots
p_n \in \{0, 1\}^n$ with $p_1 + p_2 + \cdots + p_n = m$. To see this, 
consider arbitrary row indices $i_1, i_2, \ldots, i_m \in M$ and 
column indices $j_1, j_2, \ldots, j_m \in N$. 
Note that the expanded sum contains a unique term of the 
form  $c\, a_{i_1 j_1} a_{i_2 j_2}
\cdots a_{i_m j_m}$ if and only if the indices $j_1, j_2, \ldots,j_m$ are
distinct; the coefficient $c$ is the sum of the terms $(-1)^{m - |X|}$ over
all $X \subseteq M$ that contain the row indices $i_1, i_2, \ldots, i_m$. 
If all the row indices are distinct, there is only one such set $X$, and the
coefficient correctly  equals $(-1)^{m-|M|} = 1$.  Otherwise, there are equally
many such subsets $X$ of odd and even size, and the coefficient correctly
vanishes.

To analyze the time and space complexity, we note that,  for any fixed $X
\subseteq M$, the summation over the binary sequences $p$ can be performed
using
simple dynamic programming in time $\nO(n + m(n-m))$ and space $\nO(n)$.\footnote{In the field of complex numbers,
where one can evaluate discrete convolution via fast Fourier transforms, the
time requirement can be reduced to $\nO(n \log^2 m)$. We are not aware 
whether such improvement is possible in an arbitrary (commutative) ring.} 
Here we assume that
the sets $X$ are visited in a suitable order such that each partial column sum
can be updated in an incremental fashion in constant amortized time. 
Theorem~\ref{thm:main}(iv) follows.

We end by discussing the role of commutativity. With the given definition of
permanents, Theorem~\ref{thm:main} suggests that commutativity is crucial
for efficient evaluation of permanents. However, we point out that with the
following transposed definition, the bounds in Theorem~\ref{thm:main}(ii, iv)
actually hold without
the assumption of commutativity: For an injection $\sigma$ from $M$ to $N$, denote by $\sigma_i$ the $i$th
largest element in the image $\sigma(M)$. Define the {\em transposed permanent}
of an $m \times n$ matrix $A = (a_{ij})$ over any semiring as 
\bes
	\pertp A 
	\;\doteq\; 
	\sum_{\sigma} a_{\sigma^{-1}(\sigma_1)\sigma_1}\,
	a_{\sigma^{-1}(\sigma_2) \sigma_2}\cdots
	a_{\sigma^{-1}(\sigma_m)\sigma_m}\;, 
\ees
where the summation is over all injections $\sigma$ from $\{1,2,\ldots,m\}$ to
$\{1,2,\ldots,n\}$. Note that in any commutative semiring, of course,   
$\pertp A = \per A$.

\end{document}